\documentclass[journal,12pt,onecolumn,draftcls]{IEEEtran}
\ifCLASSINFOpdf
\else
\fi

\usepackage[T1]{fontenc}
\usepackage{color}
\usepackage{graphicx}
\usepackage{epstopdf}
\usepackage{caption}
\usepackage{subcaption}
\usepackage{cite}
\usepackage{amsmath}
\usepackage{amssymb}
\usepackage{algorithmic}
\usepackage{algorithm}
\usepackage{array}
\usepackage{fixltx2e}
\usepackage{float}
\usepackage{url}
\usepackage{multirow}
\usepackage{amsthm}

\begin{document}
%
\title{Deep Reinforcement Learning in mmW-NOMA: Joint Power Allocation and Hybrid Beamforming}
%
%
%

\author{Abbas~Akbarpour-Kasgari
        and~Mehrdad~Ardebilipour
\thanks{This work has been supported financially by Iran National Science Foundation (INSF) under grant number 99011213.

A. Akbarpour-Kasgari and M. Ardebilipour are with the Department
of Electrical and Computer Engineering, K. N. Toosi University of Technology, Tehran, Iran
e-mail: mehrdad@eetd.kntu.ac.ir.}
}

%
%

\markboth{Submitted to IEEE Transactions ~~,~Vol.~ , No.~ , Date.}%
{Akbarpour-Kasgari, and Ardebilipour : DRL in mmW-NOMA: Joint power allocation and hybrid beamforming}
%



\maketitle

\begin{abstract}
High demand of data rate in the next generation of wireless communication could be ensured by Non-Orthogonal Multiple Access (NOMA) approach in the millimetre-wave (mmW) frequency band. Decreasing the interference on the other users while maintaining the bit rate via joint power allocation and beamforming is mandatory to guarantee the high demand of bit-rate. Furthermore, mmW frequency bands dictates the hybrid structure for beamforming because of the trade-off in implementation and performance, simultaneously. In this paper, joint power allocation and hybrid beamforming of mmW-NOMA systems is brought up via recent advances in machine learning and control theory approaches called Deep Reinforcement Learning (DRL). Actor-critic phenomena is exploited to measure the immediate reward and providing the new action to maximize the overall Q-value of the network. Additionally, to improve the stability of the approach, we have utilized Soft Actor-Critic (SAC) approach where overall reward and action entropy is maximized, simultaneously. The immediate reward has been defined based on the soft weighted summation of the rate of all the users. The soft weighting is based on the achieved rate and allocated power of each user. Furthermore, the channel responses between the users and base station (BS) is defined as the state of environment, while action space is involved of the digital and analog beamforming weights and allocated power to each user. The simulation results represent the superiority of the proposed approach rather than the Time-Division Multiple Access (TDMA) and Non-Line of Sight (NLOS)-NOMA in terms of sum-rate of the users. It's outperformance is caused by the joint optimization and independency of the proposed approach to the channel responses.  
\end{abstract}

\begin{IEEEkeywords}
Deep reinforcement learning, Hybrid Beamforming, millimetre-wave, Non-Orthogonal Multiple Access, Power allocation.
\end{IEEEkeywords}

%
\IEEEpeerreviewmaketitle


\section{Introduction}
%
%
%
%
\IEEEPARstart{M}{illimetre}-Wave (mmW) has two main advantages which makes it desirable in 5G networks. The first one is its abundant frequency spectrum resource, which makes it eligible to accommodate higher capacity, and the second one is the high propagation loss which makes it promising in small cell scenarios in 5G. Hence, mmW communication is the key enabler of the high data rate demands in future network generation, due to their ability to provide suitable data accommodation. 

Besides, exploiting all the mmW capacity is highly dependent on the multiple access technique which assigns different users to the high amount of spectrum resource. Among the well-known approaches, non-orthogonal multiple access (NOMA) is attracting researcher's interest to support high data rate demands and massive connectivity in 5G and beyond \cite{BenjebbourSaito, AlaviYamchi}. NOMA technique is introduced to outperform other well-known approaches such as time division multiple access (TDMA), orthogonal frequency division multiple access (OFDMA), and zero-forcing (ZF) in terms of spectral and energy efficiency \cite{ChenZhang, XuDing, DingSchober, HanifDing}. NOMA allocates frequency and time resources between all the available users, simultaneously, thanks to the power domain superposition \cite{DingYang,SaitoKishiyama, Choi, DingPeng, DaiWang, WangDai}. 

Power domain superposition is exploited by NOMA appreciating to the Successive Interference Cancellation (SIC) at the receiver. In essence, lower power is assigned to the user with better channel and more transmit power is allocated to the user with worse channel condition. Then, SIC is employed at the receiver to remove weaker users' interference. Specifically, a reasonable complexity is added at the receiver to handle the interference which is caused by non-orthogonal allocated resources \cite{CumananKrishna, DaiWang}. 

The most crucial difference of mmW-NOMA with conventional NOMA is the interlace of power allocation and beamforming problem. In \cite{DingFan}, one of pioneer works on mmW-NOMA schemes was adapted to introduce the random steering beamformer limited to the near users to each other. The energy efficiency was optimized subject to the power allocation and sub-channel assignment in \cite{FangZhang}, where no beamformer was represented. In \cite{HanifDing}, fully digital beamformer was introduced to optimize the sum-rate. Moreover, in \cite{XiaoZhu}, the joint problem was considered but the optimization was only performed on the most powerful path in channel. Further, in \cite{PangWu} joint power allocation and beamforming was optimized with considering Signal-to-Leakage-plus-Noise (SLNR) in order to decouple the power allocation and beamforming. In the contrary to all the mentioned prior works, we have used newly introduced machine learning (ML) tools to optimize the joint power allocation and beamformer design in mmW-NOMA. 

Recently, machine learning (ML) techniques are deployed in wireless communication networks as the optimization tools. Among them, Reinforcement Learning (RL) is a close-loop optimization procedure which is based on the action and reaction of agent and environment, respectively. Further, encountering a continuous problem with a large number of states will be handled with deep RL (DRL) approaches \cite{MismarEvans, LuongGagnon, WangLi, WangFeng, GeLiang}. In \cite{MismarEvans} DRL is exploited to optimize joint beamforming, power control and interference coordination in 5G network. The authors used deep Q-learning approach to estimate the future rewards. Additionally, cooperative UAV-assisted network resource allocation is optimized by DRL in \cite{LuongGagnon}. Besides, hybrid beamforming in mmW network was considered in Multi-user Multiple-Input Single-Output (MU-MISO) system utilizing multi-agent DRL-based approach in \cite{WangLi, WangFeng}. Further, DRL-based optimization for distributed dynamic MISO system was considered in \cite{GeLiang}. Employing deep Q-learning approach and cooperation of BSs, the optimal approach was designed. 

In this paper, we have exploited DRL to optimize joint power allocation and hybrid beamformer in the Base Station (BS) to increase the sum-rate of the users. The following paragraphs summarize the contributions of the work. 
\begin{itemize}
\item
In order to optimize the joint power allocation and hybrid beamforming we have modelled the joint problem based on the maximizing the sum-rate constrained to the minimum guaranteed rate for each user and the total transmission power. This problem is non-convex and could not be handled by traditional approaches, jointly. Hence, we have introduced DRL to optimize the joint problem. 
\item
In the second part, we have defined state and action spaces in the RL framework till the problem can be handled in the agent-environment interactions. We have considered the real and imaginary parts of the channel responses between the BS and users as the state space. Besides, a controlling variable $\alpha$ is defined to represent the constraints feasibility. We have considered $\alpha$ as the soft controlling variable where it is denoted based on the ratio of each constraint to the limit value of its own. Furthermore, the hybrid beamforming weights and devoted power to each user are defined as the available action space. 
\item 
Since the problem is continuous, we have utilized Soft Actor-Critic (SAC) \cite{SAC} as the suitable approach to handle the problem. SAC algorithm calculate the optimal policy by maximizing jointly the expected reward and entropy of the policy. In essence, SAC controls the policy uncertainty given the state by maximizing its entropy. This criterion makes this algorithm more robust than other well-known approaches such as Deep Deterministic Policy Gradient (DDPG). It benefits from two different DNN as actor and critic networks to approximate the action and value function, respectively. To alleviate the complexity of the problem, DNN aids the RL framework.
\item
The simulation results represent the superiority of the proposed approach regarding two different state-of-the-art approaches, called TDMA as the orthogonal multiple access (OMA) and NLOS-NOMA \cite{XiaoZhu} as the NOMA approach. Moreover, the impact of different parameters including number of antennas, signal-to-noise ratio (SNR), and minimum guaranteed rate are considered, separately. 
\end{itemize}

The rest of the paper is organized as follows: Following system model and problem formulation in section II, the proposed joint power allocation and hybrid beamformer design is represented in section III. Further, numerical results are demonstrated in section IV, and concluding remarks are presented in section V. 

\textit{Notations:}
We have used $a, \mathbf{a}, \mathbf{A}$ denoting scalar, vector, and matrix, respectively. $\mathbb{E}$ denotes the expected value of the argument. Furthermore, $\mathcal{A}, \mathcal{S}$ represent the action space, and state space, respectively. Additionally, $\mathbb{C}$ demonstrate the set of complex numbers. 




\section{System Model and Problem Formulation}
\begin{figure}
\includegraphics[width = \linewidth]{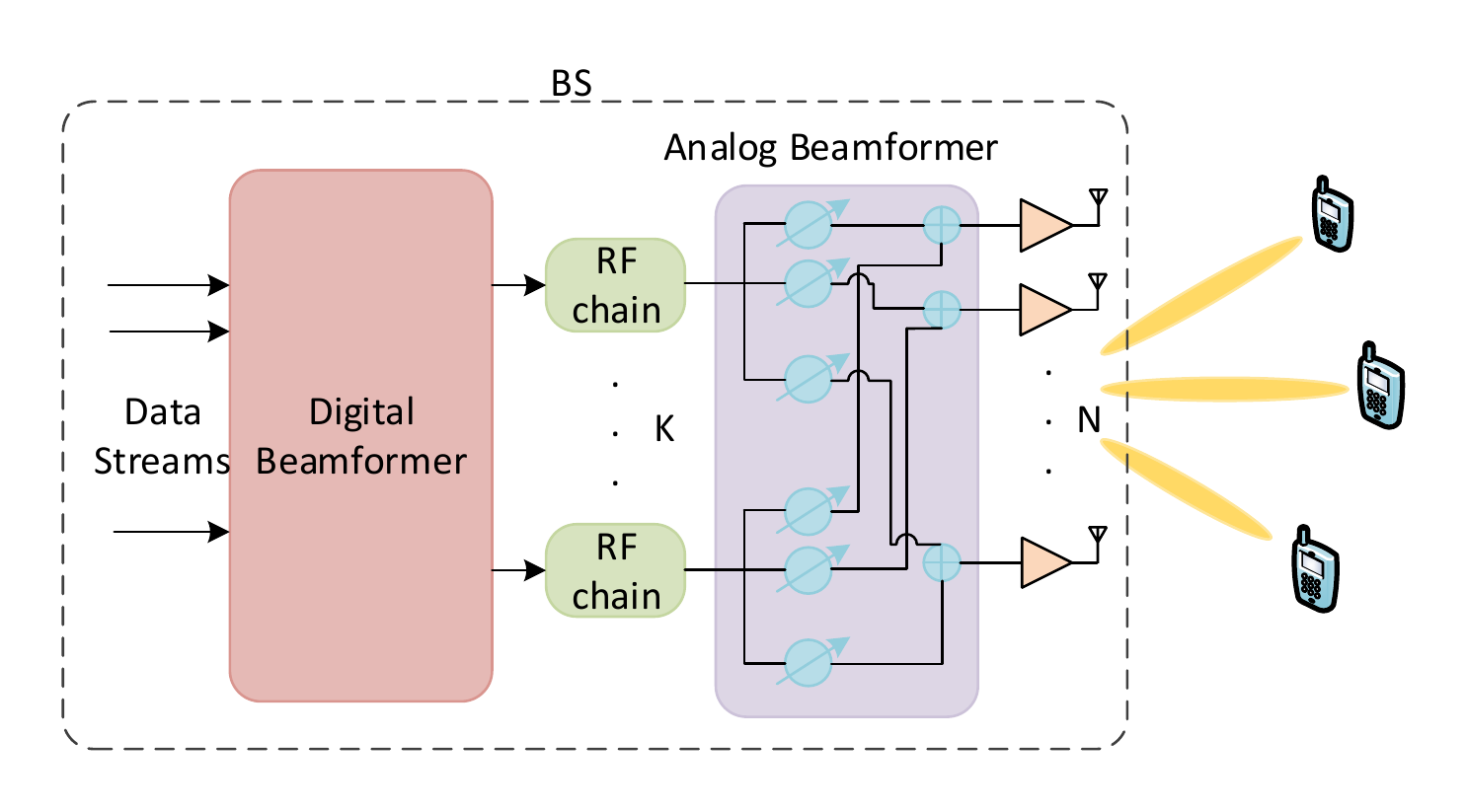}
\centering
\caption{mmW-NOMA system diagram for joint power allocation and hybrid beamforming}
\label{Fig.SystemModel}
\end{figure}
We consider a millimetre-wave non-orthogonal multiple access (mmW-NOMA) downlink system where an $N$-antenna equipped Base Station (BS) transmits towards $K$ single antenna users. Furthermore, there are $K < N$ RF chains in the BS, and $N_s$ independent data streams are transmitted from BS to the user terminals. In essence, the BS exploits $N$ antenna and NOMA strategy to simultaneously serves $K$ users. Without loss of generality, it is assumed that $K = N_s$ for touching the maximum multiplexing gain. Specifically, $N_s$ transmitted symbols are coded by digital beamformer (DBF) $\mathbf{D}\in\mathbb{C}^{K \times N_s}$ and then passed by $N$ phase shifters as analog beamformer (ABF) $\mathbf{A} \in \mathbb{C}^{N\times K}$. Besides, the ABF phase shifters are constant modulus. Hence, the overall beamformer matrix can be denoted as $\mathbf{W} = \mathbf{A} \mathbf{D}\in \mathbb{C}^{N \times N_s}$ with unit-norm columns $\mathbf{w}_k$ for $k = 1, 2, \dots, K$. Regarding NOMA as the multiple access technique, the whole power in the BS will be divided among all the users and each of the users will obtain power $p_k$ of the signal, where 
\begin{equation}
\sum\limits_{k=1}^{K}p_k = P
\end{equation}
is the total available power in the BS. Gathering the power of each user in the main diagonal of a matrix, we can represent the power denoting matrix as $\mathbf{P} = \text{diag}\{p_1, p_2, \dots, p_K\}$. Consequently, the received symbol in each user $k$ is denoted as 
\begin{equation}
y_k = \mathbf{h}_k^H \sqrt{p_k}\mathbf{w}_k s_k + \sum\limits_{\bar{k} = 1, \bar{k}\neq k}^{K}\mathbf{h}^H_{k} \sqrt{p_{\bar{k}}} \mathbf{w}_{\bar{k}} s_{\bar{k}} + n_k,
\end{equation}
where $\mathbf{h}_k \in \mathbb{C}^{N\times 1}$ represents the channel vector between the BS and the $k$-th user and $n_k$ denotes the zero-mean additive white Gaussian noise with variance $\sigma^2$ at $k$-th user terminal. Besides, $s_k$ is the transmitted symbol for $k^{\text{th}}$ user.

The channel vector between the BS and $k$-th user is an mmW channel. Since there are limited scatterers in mmW band, the multipath which is caused by reflection, is small and lead to a spatially sparse directional channel in angle domain \cite{PengWang,GaoHu}. Assuming a uniform linear array (ULA), the mmW channel can be demonstrated as \cite{PengWang,GaoHu}
\begin{equation}
\mathbf{h}_k = \sum\limits_{l = 1}^{L_k}\lambda_{k,l}\mathbf{a}(N, \Omega_{k,l})
\end{equation}
where $\lambda_{k,l}$, and $\Omega_{k,l}$ are the complex coefficient and angle-of-arrival (AoD) of the $l$-th multipath components in the channel vector for the $k$-th user, respectively. Further, $L_k$ is the number of multipath components for $k$-th user, and $\mathbf{a}(.)$ is the steering vector function which is defined as 
\begin{equation}
\mathbf{a}(N, \Omega) = \left[ e^{j\pi 0 \cos(\Omega)}, e^{j\pi 1 \cos(\Omega)}, \dots, e^{j\pi (N-1) \cos(\Omega)}  \right]
\end{equation}
and is dependent to the geometry of the array. 

For the rest of the paper, we assume that the user terminals are ordered based on their channel quality, i.e., $\| \mathbf{h}_1 \|_2 \leq \| \mathbf{h}_2 \|_2 \leq \dots \leq \| \mathbf{h}_K \|_2$. Exploiting Successive Interference Cancellation (SIC), the NOMA receiver can decode the power domain transmitted signal in the same frequency and time resource \cite{WunderJung, ZhangHanzo}. Based on the suggested ordering and exploiting SIC in the receiver, each user $k$ can remove first $k-1$ signals as interference and treats other remaining users, i.e., from $k+1$ to $K$, as noise \cite{BenjebbourSaito, HanifDing}. Obviously, this kind of detection is not optimal, but it is more than our paper discussion and here we try to demonstrate a model-free approach for optimization. 
Therefore, the received Signal-to-Interference-plus-Noise (SINR) in each user can be defined as 
\begin{equation}
\text{SINR}_k = \frac{|\mathbf{h}^H_k \mathbf{w}_k|^2 p_k}{\sum\limits_{k^\star = k + 1}^{K}|\mathbf{h}^H_k \mathbf{w}_{k^\star}|^2 p_{k^\star} + \sigma^2} 
\end{equation}
where $k^\star$ indicates the user indices more than the current index. Hence, the achievable rate in each user is denoted as 
\begin{eqnarray}
R_k = \log_2(1 + \text{SINR}_k). 
\end{eqnarray}

In this paper, we consider the problem of joint power allocation and hybrid beamformer design for sum-rate optimization. As a consequence, the objective function is the sum-rate of the $K$ users. Moreover, there are some constraints which are denoted as below: 

\begin{align} \label{eq.opt} \nonumber
 \max_{\mathbf{d}_k,\mathbf{a}_k,p_k} & \sum\limits_{k=1}^{K}R_k\\ \nonumber 
 \text{s.t.      } & R_k \geq r_k \\ \nonumber
 & \sum\limits_{k=1}^{K}p_k = P \\ \nonumber
 & \| \mathbf{w}_k \|_2^2 = 1 \\ 
 &  |A_{i,j}| = 1/\sqrt{N} 
\end{align}

where $k = 1,2, \dots, K$, $i = 1,2,\dots, N$, and $j = 1,2,\dots, K$ for the above equation and $P$ is the total available power in BS. Besides, $\mathbf{d}_k$, and $\mathbf{a}_k$ denote the digital and analog beamformer weights for the $k^{th}$ user, respectively. Moreover, $r_k$ is the minimum guaranteed rate in $k$-th user which should be provided. As discussed earlier, the cost function is the sum-rate of the users. Further, the first constraint ensures the minimum rate for each user. While, the second and third constraints determine the feasible space of the independent variables of the problem, i.e., they guarantee the transmission power of the users and their corresponding beamforming weights. Ultimately, the last one restricts the analog beamformer behaviour. 

The signal decoding in SIC-based receivers is performed after decoding of weaker users' signal and successfully removing their corresponding interference. In order to aid this kind of decoding, the received power of the desired signal should be higher than other users' power level. This phenomena will lead to some equations on the signal power which imposes more power to the users located far from BS \cite{AlaviCumanan}. 



\section{Joint Power Allocation and Hybrid Beamformer Design}
Here we will discuss the proposed approach to solve the joint power optimization and hybrid beamformer design problem in Eq. \eqref{eq.opt}. We have utilized the Deep Reinforcement Learning (DRL) approach to introduce the optimized power and beamformer weights. 

\subsection{Deep Reinforcement Learning}
Deep Reinforcement Learning (DRL) is involved of Deep Learning (DL) and Reinforcement Learning (RL), simultaneously. Due to existence of the DL part in this approach, it's very suitable handling large state and action space MDP problems \cite{Sutton}. The MDP can be defined by a four-tuple space as $\{\mathcal{S}, \mathcal{A}, \mathcal{P}, \mathcal{R}\}$, where $\mathcal{S}$ is the state space denoting the observation from environment. Further, $\mathcal{A}$ is the action space and representing the available actions from the agent to optimize the reward. Besides, $\mathcal{P}$ is the state transition probability, and $\mathcal{R}$ is the immediate reward of the action $a_t\in \mathcal{A}$ in state $s_t \in \mathcal{S}$. According to the immediate reward, the long-term reward 
\begin{equation}
V(s) = \mathbb{E}\lbrace \sum\limits_{t=0}^{\infty}\gamma^t r_t(s_t,a_t) | s\rbrace
\end{equation}
for discount factor $\gamma \in [0,1]$ can be maximized to guarantee the best action in each state. 

\begin{figure*}
\includegraphics[width = \textwidth]{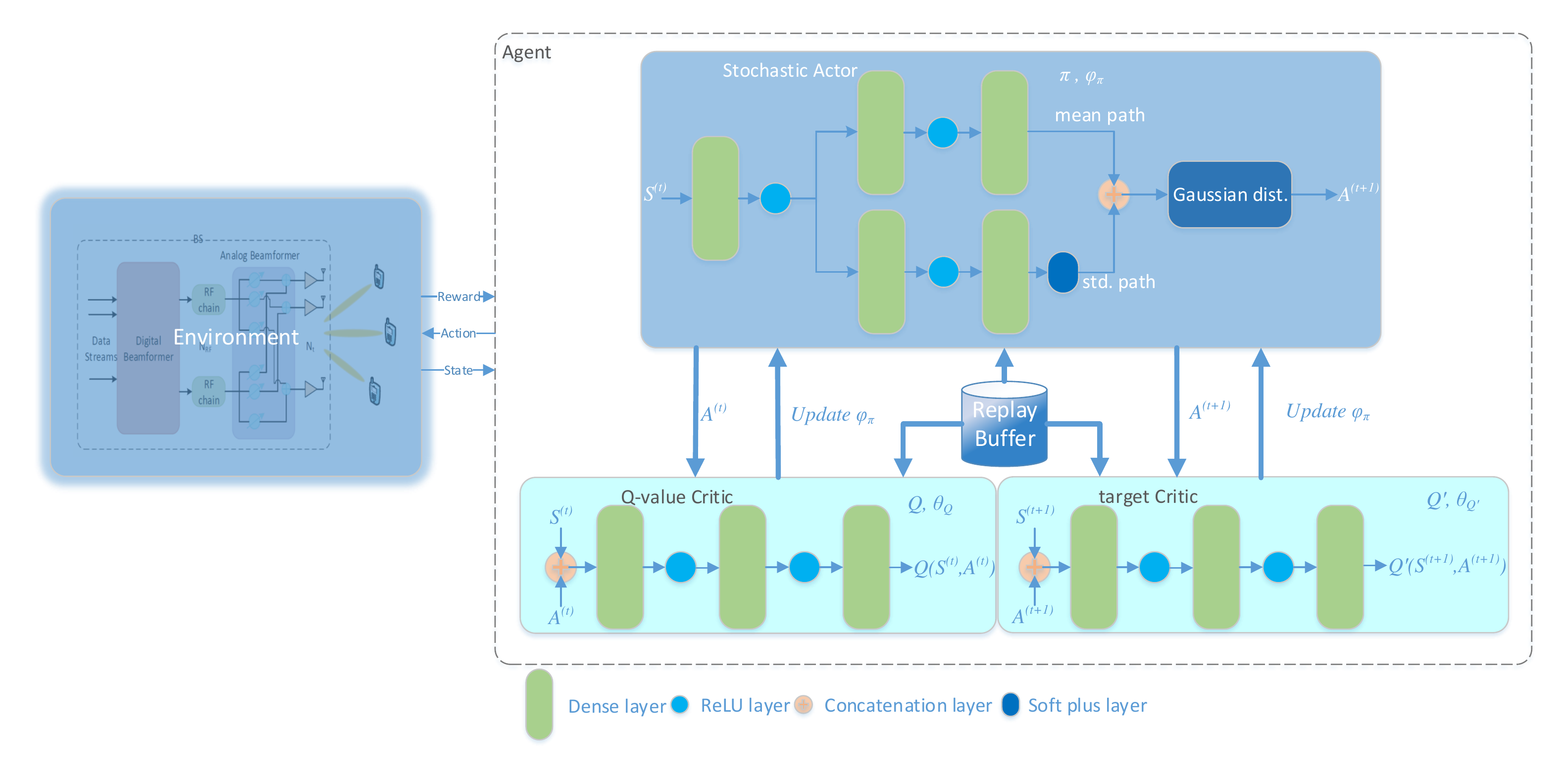}
\centering
\caption{Schematic diagram of soft actor-critic network}
\label{Fig.ActorCriticNet}
\end{figure*}

Accordingly, we can model the problem at hand into DRL framework by defining mandatory 4-tuple state variables. In order to represent the 4-tuple state variables, we demonstrate the immediate reward function based on the problem. To this end, we need to reconsider the optimization problem in Eq. \eqref{eq.opt}. The auxiliary variable $\alpha$ is defined to guarantee the constraints feasibility. Hence, the following immediate reward is defined 
\begin{equation}
	r_t = \alpha R
\end{equation}
where $R = \sum_{k=1}^{K} R_k$, and $\alpha = \prod\limits_{i=1}^{K+1} \alpha_i$ is the controlling parameter for guaranteeing the constraints of Eq. \eqref{eq.opt}. By the following definition for $\alpha_i$
\begin{equation} \label{eq.a1}
\alpha_i = \left \lbrace 
\begin{aligned}[l|l]
\frac{R_k}{r_k}, && \text{if } R_k < r_k \\
1, 		  && \text{otherwise}
\end{aligned}
\right.
\end{equation}
for $1 \leq i \leq K$, and
\begin{equation} \label{eq.a2}
\alpha_i = \left \lbrace 
\begin{aligned}[l|l]
\frac{P}{\sum\limits_{k=1}^{K}p_k} , && \text{if } P < \sum\limits_{k=1}^{K}p_k \\
1, 		  && \text{otherwise}
\end{aligned}
\right.
\end{equation}
for $i = K + 1$. 
Apparently, Eq. \eqref{eq.a1} ensures the minimum guaranteed rate of $K$ users, while the Eq. \eqref{eq.a2} guarantees the power limitation constraint in Eq. \eqref{eq.opt}. With the aid of $\alpha$ we are able to weight the received sum-rate of user based on the feasibility of the constraints. In essence, whenever $\alpha_i$ for $i = 1,2, \dots, K + 1$ are $1$ it determines that the designed power allocation and hybrid beamformer strategy satisfies the constraints, though, the calculated rate is not downgraded. While each of the $\alpha_i$ for $i = 1, 2,\dots, K+1$ is the calculated ratio in Eq. \eqref{eq.a1} and \eqref{eq.a2}, it would decrease the sum-rate relative to the calculated ratio. 


Additionally, the state $\mathcal{S}$ is defined as $\mathcal{S} = \lbrace \mathbf{h}_1^{\mathcal{R}}, \mathbf{h}_1^{\mathcal{I}}, \mathbf{h}_2^{\mathcal{R}}, \mathbf{h}_2^{\mathcal{I}},  \dots, \mathbf{h}_K^{\mathcal{R}}, \mathbf{h}_K^{\mathcal{I}}, \alpha \rbrace$, where superscript $\mathcal{R}$ and $\mathcal{I}$ denote real and imaginary parts, respectively. Furthermore, the action space which includes the available actions that are taken by the agent, can be denoted by $\mathcal{A}$ and is defined as $\mathcal{A} = \lbrace \mathbf{d}_1^{\mathcal{R}}, \mathbf{d}_1^{\mathcal{I}}, \mathbf{a}_1^{\mathcal{R}}, \mathbf{a}_1^{\mathcal{I}}, \dots, \mathbf{d}_K^{\mathcal{R}}, \mathbf{d}_K^{\mathcal{I}}, \mathbf{a}_K^{\mathcal{R}}, \mathbf{a}_K^{\mathcal{I}}, p_1, p_2, \dots, p_K \rbrace$, where $\mathbf{d}_k$, and $\mathbf{a}_k$ are the digital and analog beamformer weights for the $k^{th}$ user, respectively. 
\subsection{Soft Actor-Critic}
DNNs are employed by the researchers to enable self-decision making in RL framework. But for real-world problems there are two crucial challenges: sample complexity, and hyperparameter sensitivity. Sample complexity is regard as the millions of samples that these approaches needed to model an even simple task. On the other hand, achieving suitable results in each problem is highly dependent to the learning rates, exploration constants and other hyperparameter settings. Soft Actor-Critic (SAC) formulation represents significant improvement in robustness and exploration \cite{SAC}. These two essential characteristics are represented in \cite{Ziebart, HaarnojaTang}. SAC algorithm combines three pivotal components: firstly, an actor-critic framework with individual networks for policy and value function, secondly, an off-policy criterion that enables the reusability of previous data collection, and finally, entropy maximization which brings stability and exploration to the algorithm. SAC incorporates off-policy actor-critic training with a stochastic actor and targets to maximize the entropy of the stochastic actor \cite{SAC}. 

As discussed earlier, continuous problems need DNN as model-free policy and value estimators, separately. Parametrized Q-function approximator is $Q_\theta(s_t, a_t)$ and a tractable policy is denoted by $\pi_\phi(a_t | s_t)$. In these notations, $\phi$ and $\theta$ are actor and critic DNN parameters, respectively. Accordingly, the critic network will be trained to minimize the soft Bellman residual as
\begin{eqnarray}
J_Q(\theta)& = &\mathbb{E}_{(s_t, a_t)}\left[ (Q_\theta(s_t,a_t) - P_{t+1}(s_t,a_t))^2 \right] \\
J_\pi(\phi)& = &\mathbb{E}_{s_t}\left[\mathbb{E}_{a_t}\left[\alpha\log(\pi_\phi(a_t|s_t))-Q_\theta(s_t,a_t)\right]\right]
\end{eqnarray}
while 
\begin{equation}
P_{t+1}(s_t,a_t) = r(s_t,a_t) + \gamma\mathbb{E}_{s_{t+1}}\left[ V_{\bar{\theta}}(s_{t+1}) \right]
\end{equation}
where $V_{\bar{\theta}}(s_{t+1})$ is the value function regarding the next state, and $r(s_t,a_t)$ denotes the reward function in the state $s_t$ taking action $a_t$. 
$J_Q(\theta)$ can be optimized by stochastic gradients. While, in the update process target soft Q-function is exploited which defined as exponentially moving average of previous weights leading to stabilization in training \cite{MnihKavukcuoglu}. Besides, $J_\pi(\phi)$ is expected weighted Kullback-Leibler (KL)-divergence where $a_t = f_\phi(\epsilon_t,s_t)$ represents the neural network approximation for policy, and can be optimized by gradient-based approaches leading to any tractable stochastic policy \cite{SAC}. 
The whole network of the system is demonstrated as in Fig. \ref{Fig.ActorCriticNet}.


\subsection{Deep Neural Networks}
As mentioned earlier, the whole optimization network involved of two independent DNNs. The first one which is the critic one, accepts three different inputs including current observation, current action, and current reward. The first two ones (i.e., observation and action) are merged via a concatenation layer and then go across two dense layers followed by ReLU as the activation layer. Finally, a dense layer will estimate the Q-value of the network. Correspondingly, there is another critic network called target critic while will be updated periodically based on the latest parameters values and critic in order to improve the stability. 

Secondly, the actor network welcomes the current observation and after applying a dense layers with ReLU as the activation layer, it will be break into two separate paths, called mean path and standard deviation (std) path to represent the Gaussian probability density function (pdf). In the mean path, after applying a dense layer and ReLU activation, there is a dense layer and mean value will be estimated. While, in the std path, after employing a dense layer and ReLU activation, there is the final dense and a soft plus activation layer to estimate the std value of the pdf. Then, the agent will generate the desired action randomly, based on the estimated Gaussian pdf.

\section{Numerical Results}
In this section, the numerical results of the proposed approach is compared with other approaches in order to demonstrate its ability for optimizing the problem. This section is consist of two separate parts. In the first part, the learning curve of the proposed SAC-based approach is demonstrated and compared with DDPG approach as the benchmark of continuous problem dealing approach. In the second part, the proposed sum-rate is represented and compared with 'TDMA' and 'NLOS-NOMA' in \cite{XiaoZhu} as state-of-the-art approaches for sum-rate problem. The common parameters for learning system are listed in Table \ref{table.1}. 

\begin{table}
\centering
\caption{Simulation parameters for DRL}
\begin{tabular}{| r | l |}
\hline
Parameter & Value \\
\hline \hline
Discount factor & $\gamma = 0.99$ \\
Episode trial & $100$ \\
Moving average length & $25$ \\
Actor learn rate & $\alpha_a = 10^{-3}$ \\
Critic learn rate & $\alpha_c = 10^{-3}$ \\ 
Experience buffer length & $10^{6}$ \\
Target smooth factor & $\tau = 10^{-3}$ \\
\hline
\end{tabular}
\label{table.1}
\end{table}

\subsection{Learning curve}
\begin{figure}
	\includegraphics[width=0.99\linewidth]{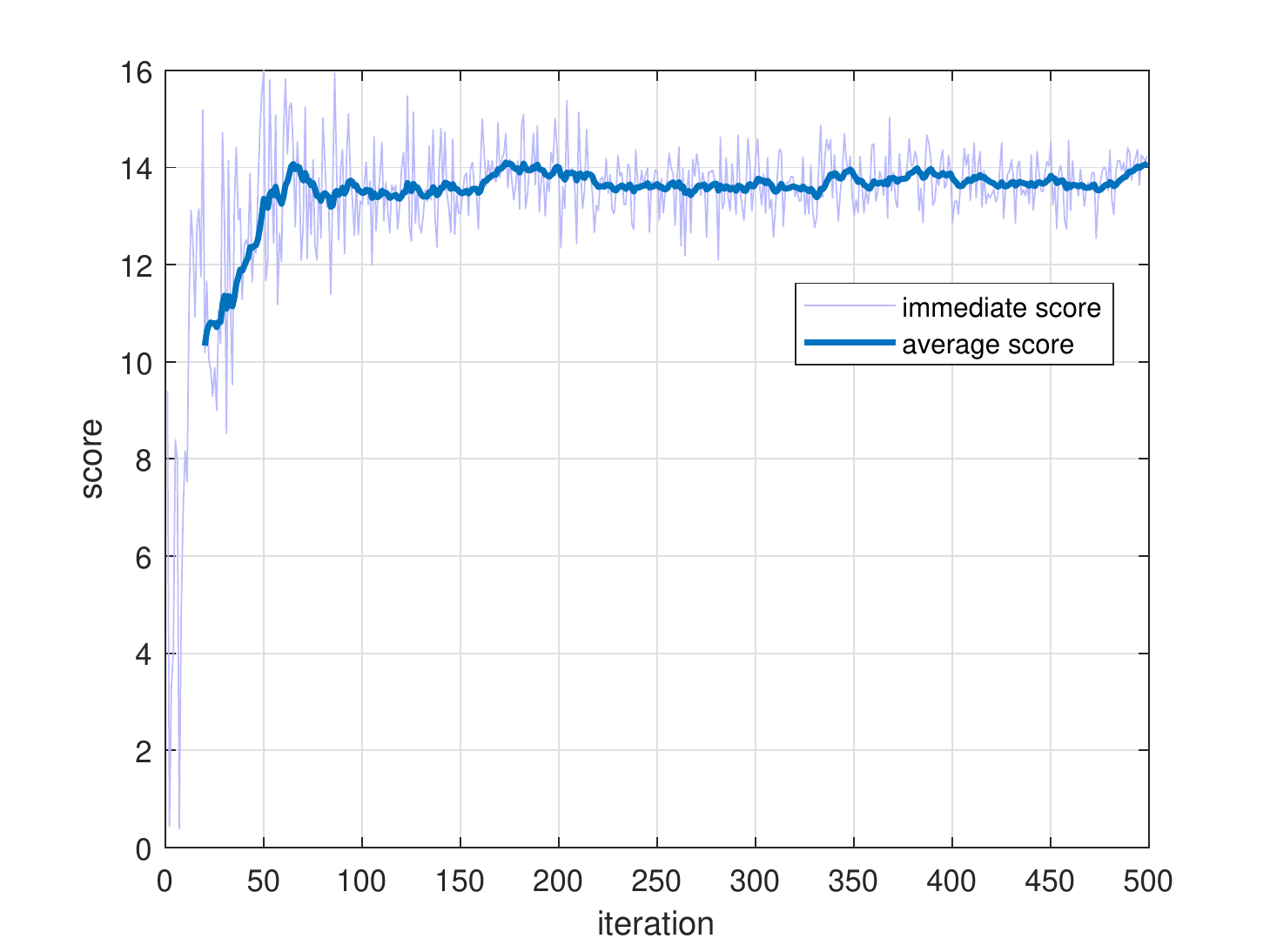}
    \caption{The immediate and average score of the SAC algorithm}\label{fig.score}
\end{figure}

As mentioned earlier, the learning curve of the SAC-based algorithm is represented in this section. The first simulation is the comparison of average and immediate reward for $K=2$ user system with a BS equipped with $N = 32$ antennas, and minimum achievable rate of $r_1 = r_2 = 3$ (bps/Hz) is depicted in Fig. \ref{fig.score}. The average is computed on the last $25$ scores. As can be seen, the average reward is reached over $13$ (bps/Hz) after nearly $40$ episodes. 

\begin{figure}
	\includegraphics[width=0.99\linewidth]{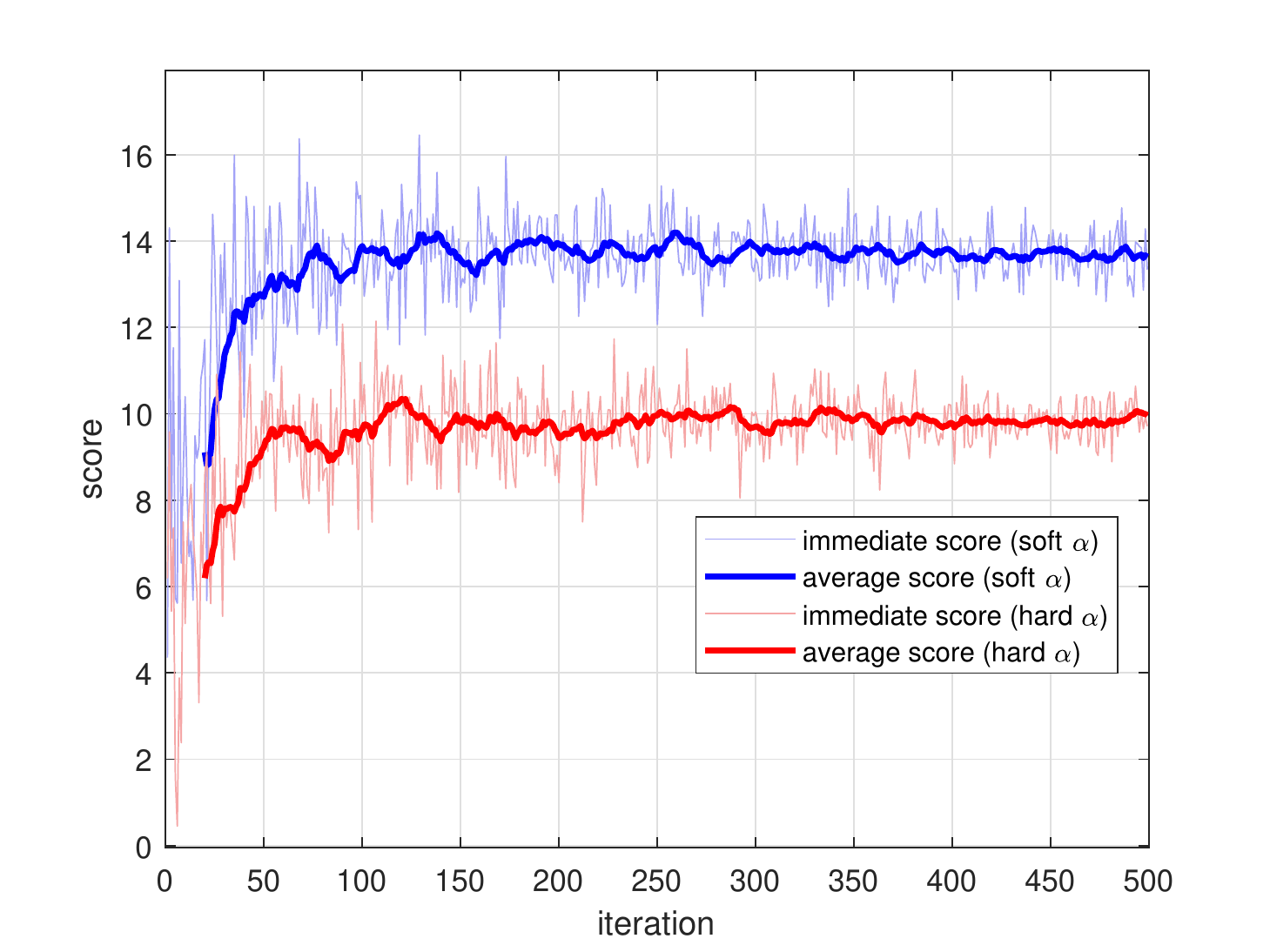}
	  \caption{The impact of hard or soft $\alpha$ on the immediate and average score}\label{fig.alpha}
\end{figure}

In Fig. \ref{fig.alpha}, the impact of soft and hard $\alpha$ are compared. The system configuration is as before. Soft $\alpha$ is presented in this paper, earlier in Eq. \eqref{eq.a1} and Eq. \eqref{eq.a2}. Hard $\alpha$ is the case where $\alpha$ is a binary value which is $1$ in case of all the constraints feasibility and is $0$ in case of infeasibility for at least one of the constraints. Since, in hard $\alpha$ the immediate reward is zero and we cut the value of the reward with a hard limiter, the overall immediate and average reward is lower than the case of soft $\alpha$ almost $3$ (bps/Hz). 

\begin{figure}[htb]
\includegraphics[width=0.99\linewidth]{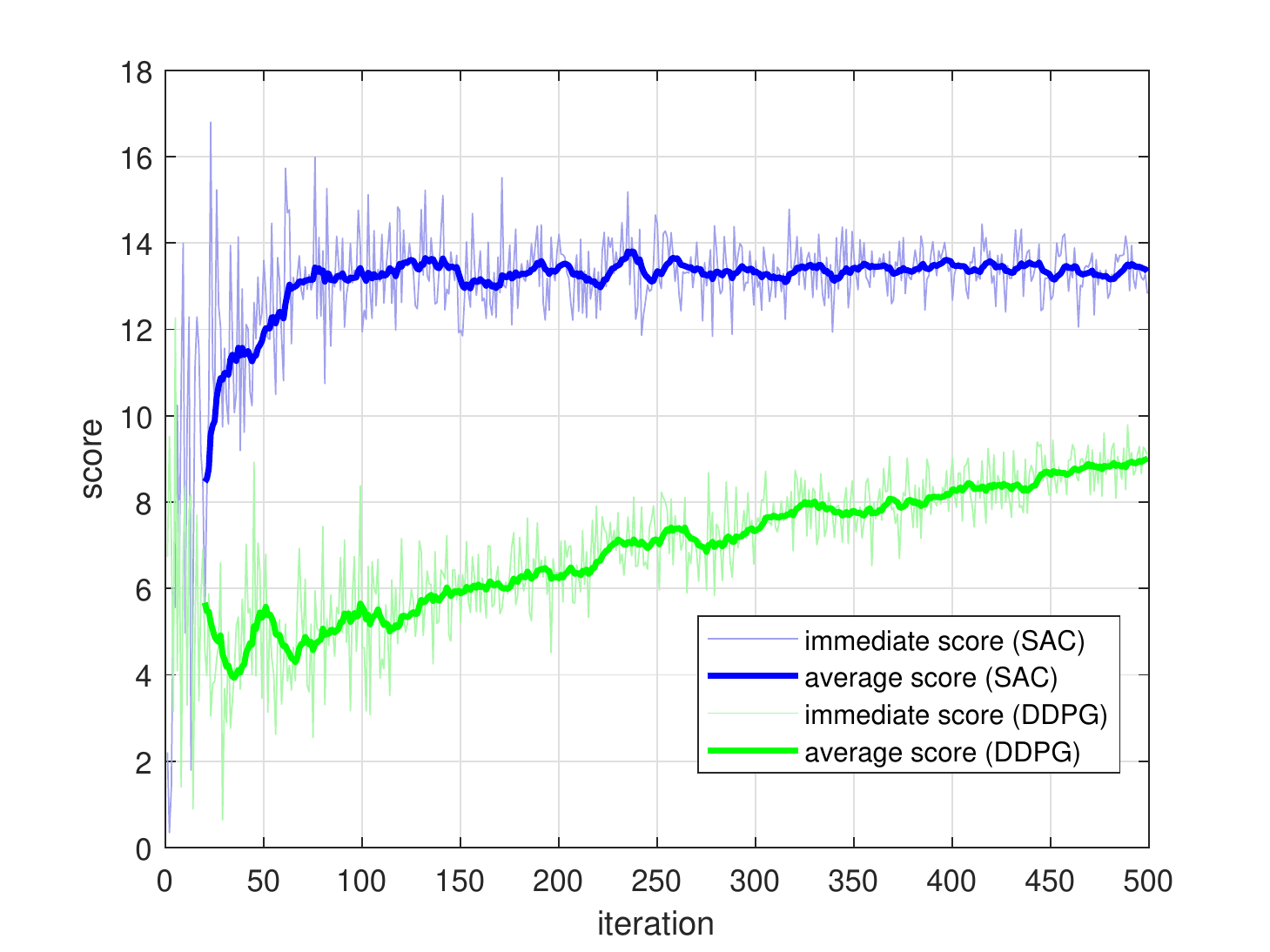}
  \caption{The comparison of DDPG and SAC based DRL approaches}\label{fig.algorithm}
\end{figure}

Finally, the comparison of the proposed SAC-based algorithm with the DDPG-based approach is presented in Fig. \ref{fig.algorithm}. Again, the configuration of simulation is the same as previous ones. Obviously, the speed of convergence in DDPG is dominated by SAC-based algorithm. Since the entropy of policy and reward are optimized jointly, the convergence speed and stability is increased drastically. 

\subsection{Sum-rate}
Here, we will consider the optimization capability of the proposed DRL-based approach rather than 'TDMA' which is an instance of Orthogonal Multiple Access (OMA) approach and 'NLOS-NOMA' which is an instance of NOMA-based approach and is designed in \cite{XiaoZhu}. We have regarded these two approaches as the state-of-the-art methods. Additionally, we have considered four different parameter settings to represent the capability of the apprach in all the conditions. Hence, the results of the simulations are represented in four different figures in the following paragraphs. In order to compare the results, we have trained the network with $10^6$ independent samples during $1000$ consecutive episodes. Then, the results are calculated based on $250000$ new samples, i.e., the results are averaged over $250000$ independent runs. 

\begin{figure}[t!]
	\includegraphics[width=0.99\linewidth]{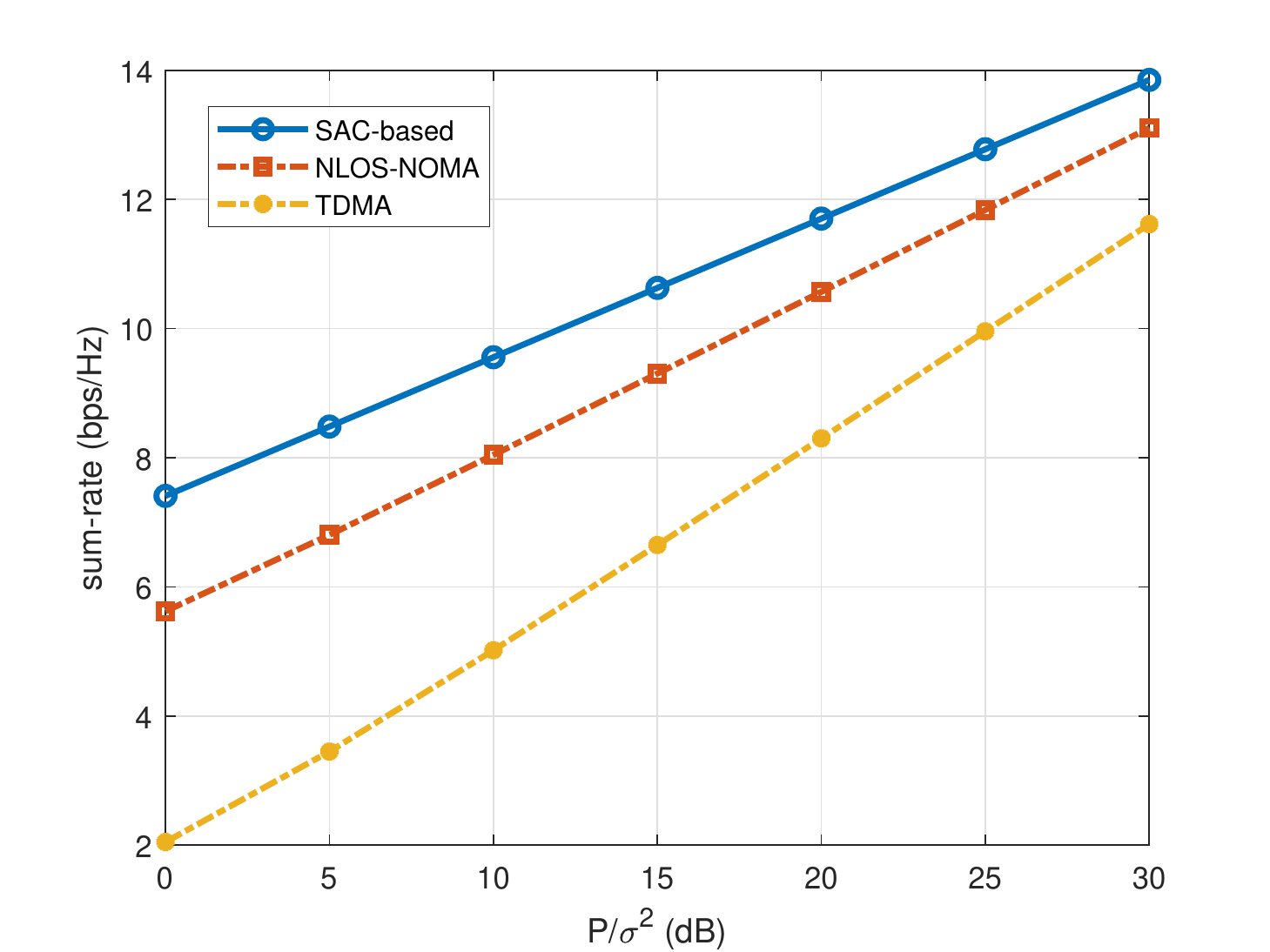}
    \caption{Sum-rate comparison for various SNRs}\label{fig.snr}
\end{figure}

As the first one, the Fig. \ref{fig.snr} demonstrated the sum-rate performance of the approach regarding the received signal-to-noise ratio (SNR) of the users. In this simulation, $K = 2$ users are served by a $16$-antenna BS while the minimum guaranteed rates are $r_1 = r_2 = 3 $(bps/Hz).  Moreover, SNR is defined as the ratio of the received power $P$ to the noise variance $\sigma^2$ which is $\sigma^2 = 1 $(mW). As depicted in Fig. \ref{fig.snr}, the proposed approach outperforms  the other two methods. The superiority of the proposed approach rather than the 'TDMA'-based approach is obvious (due to the resource sharing in NOMA systems). In addition, the proposed approach is superior to the 'NLOS-NOMA' approach which is caused by the joint optimization of the proposed approach rather than the successive manner of 'NLOS-NOMA' approach. Furthermore, 'NLOS-NOMA' is highly dependent to the condition of the channel while our proposed approach is independent of the channel condition. 

In the next simlation, the proposed approach is compared with the others based on the user numbers which are served by the BS. For this simulation, we have considered $30$ dB SNR for the users. Furthermore, $N_{ant} =16, r_i=3$ for $i = 1,2,\dots,8$. Expectedly, the sum-rate is decreased with the increase of the number of users. It is resulted from the increased interference coming from the users. Again, the proposed approach outperforms the 'NLOS-NOMA' based approach because of channel insensitivity.

\begin{figure}
	\includegraphics[width=0.99\linewidth]{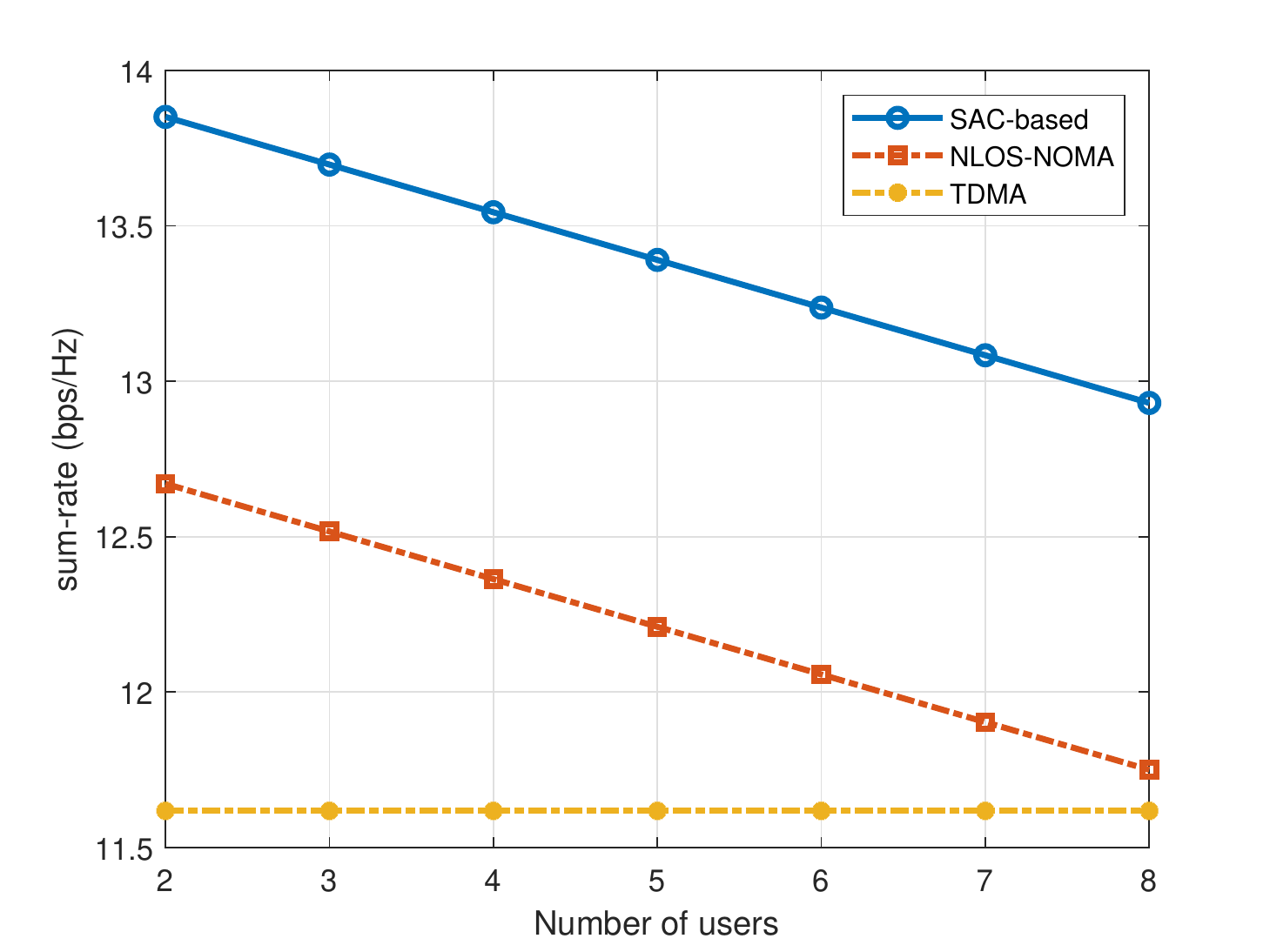}
	  \caption{Sum-rate comparison for different number of users}\label{fig.user}
\end{figure}

\begin{figure}
\includegraphics[width=0.99\linewidth]{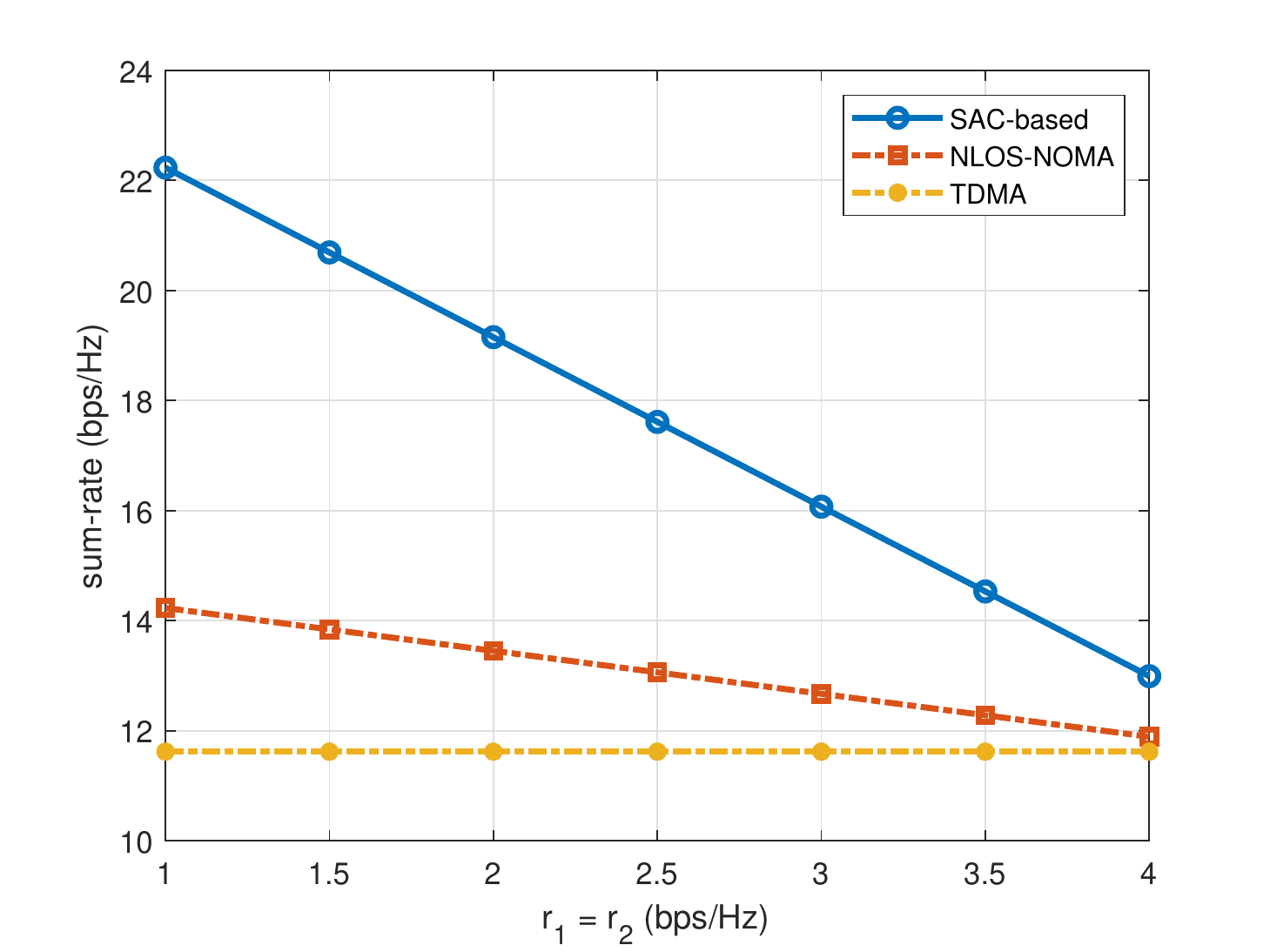}
  \caption{Sum-rate comparison for minimum guaranteed rate in users}\label{fig.minRate}
\end{figure}

The impact of the minimum guaranteed rate on the sum-rate of the approaches is regarded in Fig. \ref{fig.minRate}. Here, two users are supported with a $16$-antenna-equipped BS.  The minimum guaranteed rate is changed from $1$ to $4$ with steps equal to $0.5$. As represented in Fig. \ref{fig.minRate}, the sum-rate is decreased with the increase of the minimum guaranteed rate. It is caused by the increase of the power which is devoted to the bad-conditioned user in the system to maintain the minimum rate. Hence, lower power remains for the other user. This is the main reason of the decrease of the sum-rate with the increase of the minimum guaranteed rate. 

\begin{figure}
\includegraphics[width=0.99\linewidth]{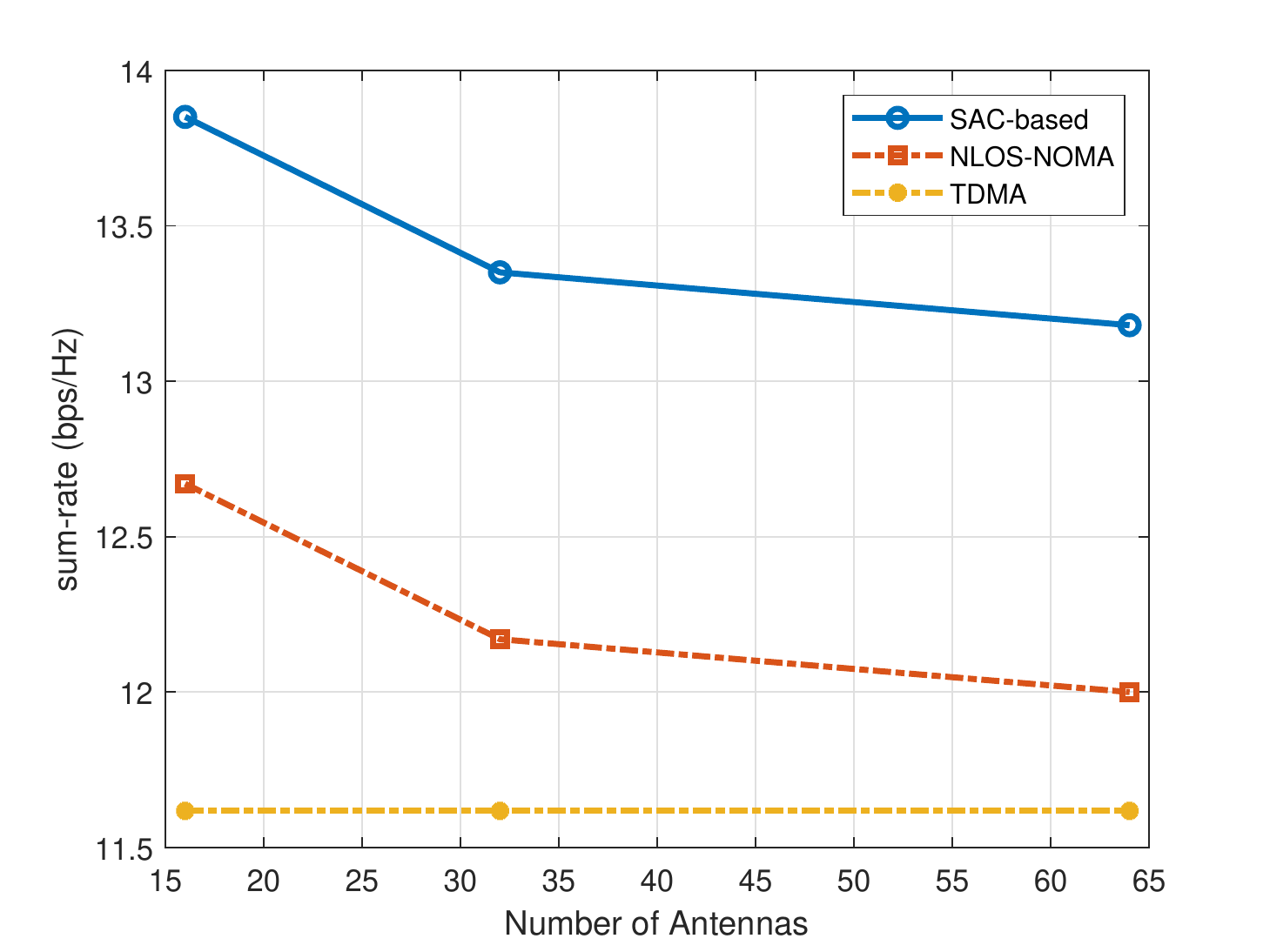}
  \caption{Sum-rate comparison for several number of antennas}\label{fig.antenna}
\end{figure}

Finally, antennas impact is represented in Fig. \ref{fig.antenna}. As depicted the increase of the number of antenna, decrease smoothly the sum-rate of the two users with $30$-dB SNR. The antenna numbers are $16, 32, 64$. Increase of the antennas will increase the interference which generates lower sum-rate. While in TDMA case, the interference is omitted because of the orthogonal access to the resources. 

\section{Concluding Remarks}
In this paper, we have considered the design of power allocation and hybrid beamforming strategies, jointly. We have exploited control-theory-borrowed DRL approach to optimize the sum-rate of the users in mmW-NOMA based system. We have modelled the optimization problem in the context of the DRL, and due to the continuity of the problem, we have utilized SAC-based approach. The system is included of two DL framework act as the critic and actor. To optimize the joint power allocation and hybrid beamforming we have defined a soft sum-rate reward which is based on the received rate of each user and consumed power. In the simulations, the proposed approach outperforms other benchmark approaches. There are two main reasons for the superiority of the proposed approach as the joint optimization behaviour, and independence to the channel models. The joint optimization behaviour helps the BS to design a power allocation and hybrid beamforming strategies which cooperate with each other to optimize the sum-rate, while in other successive approaches the optimal beamformer is designed regarding the power allocated strategy, and again power allocated strategy is designed based on the selected beamformer. In this work, channel estimation error is not considered which will be handled in our future works. 

\ifCLASSOPTIONcaptionsoff
  \newpage
\fi

\end{document}